\documentclass[11pt]{cernrep}
\usepackage{graphicx,epsfig,amsmath}
\usepackage{subcaption}
\usepackage{hyperref}
\usepackage{cite,color}
\usepackage{siunitx}

\newcommand{\bbbar}{\ensuremath{\bar{b}b}}
\newcommand*{\TeV}{\ensuremath{\text{Te\kern -0.1em V}}}
\newcommand*{\GeV}{\ensuremath{\text{Ge\kern -0.1em V}}}
\newcommand{\pt}{\ensuremath{p_{\mathrm T}}}

\begin{document}

\title{DM+$b\bar b$ simulations with {\sc DMSimp}: an update}
\author{Y.~Afik$^1$, F.~Maltoni$^{2,3}$, K.~Mawatari$^{4}$, P.~Pani$^{5}$, G.~Polesello$^{6}$, Y.~Rozen$^{1}$, M.~Zaro$^{7}$}
\institute{
$ $
\\$^1$ Department of Physics, Technion: Israel Institute of Technology, Haifa, Israel
\\$^2$ Centre for Cosmology, Particle Physics and Phenomenology (CP3),
Universit\'e catholique de Louvain, B-1348 Louvain-la-Neuve, Belgium
\\$^3$ Dipartimento di Fisica e Astronomia, Universit\`a di Bologna and INFN, Sezione di Bologna, via Irnerio 46, 40126 Bologna, Italy
\\$^4$ Department of Physics, Osaka University, Toyonaka, Osaka 560-0043, Japan
\\$^5$ Deutsches Elektronen-Synchrotron DESY, Hamburg and Zeuthen, Germany
\\$^6$ INFN, Sezione di Pavia, Via Bassi 6, 27100 Pavia, Italy
\\$^7$ Nikhef, Science Park 105, NL-1098 XG Amsterdam, The Netherlands}
\maketitle

\begin{abstract}
Searches for dark matter (DM) produced using collider data probe wide regions of the
allowed parameter space of many models and have become competitive with more traditional
searches. The interpretations of the results can be efficiently performed  in simplified models, which feature only a mediator and a DM candidate together with the interactions among them and the standard model particles. The {\sc DMSimp} model in {\sc FeynRules} currently features a wide set of $s$-channel simplified models and can be exploited by the {\sc MadGraph5\_aMC@NLO} framework to produce event samples  including NLO QCD corrections for realistic simulations. Higher-order corrections typically have a sizeable impact on the total production rate and lead to a reduction of the theoretical uncertainties. In this short note we report on a recent update of the {\sc DMSimp} model, which makes it possible to simulate dark matter production in association with bottom quarks in a mixed four-flavour scheme, where  the $\overline{\rm MS}$ renormalisation for the bottom-quark Yukawa is employed, while the bottom quark mass is on shell.  By comparing with five-flavour scheme, we show that the mixed four-flavor scheme 
 provides reliable predictions for  DM+$b\bar b$ final states in a wide range of DM masses.
\end{abstract}

\section{Introduction}

The most recent searches for Dark Matter (DM) performed by the ATLAS and CMS collaborations rely on a framework of simplified models  characterised by one new particle that mediates the interaction between the dark sector and the Standard Model (SM). These simplified models have been systematically categorised in the
DM Forum Report \cite{Abercrombie:2015wmb} in terms of parameter space and collider
signature. In LHC Run-1 and especially in Run-2, the ATLAS and CMS
collaborations have examined a large variety of these models and
signatures \cite{Aaboud:2017rzf,Aaboud:2017aeu,ATLAS:2016jaa,ATLAS:2016ljb,ATLAS:2016xcm,ATLAS:2016tsc,Sirunyan:2017xgm,CMS:2016mxc,CMS:2016jxd,CMS:2016uxr}. 

Motivated by the significant impact of these searches and the corresponding interpretations in the context of simplified models, a very general implementation of these models in the
{\sc FeynRules}~\cite{Alloul:2013bka}/{\sc MadGraph5\_aMC@NLO} \cite{Alwall:2014hca}
({\sc MG5aMC} henceforth) framework, dubbed {\sc DMSimp}, accurate up to NLO in QCD has been released~\cite{Mattelaer:2015haa,Backovic:2015soa,Neubert:2015fka}. It features $s$-channel mediators (SM gauge singlet, spin 0 or 1 bosons) coupling to DM and SM particles.
Predictions and event generation for this class of models can be achieved at NLO QCD accuracy, {\it in a fully automatic way}, for a wide set of observable/final states, while also employing matching/merging multi-jet techniques~\cite{Frederix:2012ps}. 

In this short note, we consider the specific case of $s$-channel spin-0 mediators produced in association with heavy quarks (DM+HF), for which the sensitivity with Run-2 data started to probe an interesting
phase space of parameters \cite{Aaboud:2017rzf,Aaboud:2017aeu,Sirunyan:2017xgm}. 
It was already shown in the literature that higher-order QCD corrections provide a sizeable impact on production
rates and reduction of the theoretical uncertainty for the DM production associated with a top-quark pair \cite{Backovic:2015soa}. We focus on the production of DM particles in association with bottom quarks (denoted DM+\bbbar) 
and in particular on the issue of choosing the most appropriate scheme (which might depend on the mass of the mediator) for the simulation of final states involving missing transverse momentum and one or two $b$-jets.  As for the case for Higgs+\bbbar\ production, which has been widely discussed in the literature, e.g.,~\cite{deFlorian:2016spz}, a key ingredient for improving the convergence of the perturbative series and therefore the reliability of the predictions, is the choice of the flavour scheme and of  the renormalisation of the bottom-quark coupling with the scalar mediator (which we will dub bottom Yukawa in the following). The aim of this note is to explicitly show that also for the DM case, the choice of a mixed scheme (on-shell for the bottom-quark mass and $\overline{\rm MS}$ for the bottom-Yukawa coupling) provides reliable predictions for a wide range of mediator masses and signatures.

\section{Details of the simulation}
\label{sec:technicalities}

As it is well known,  processes featuring $b$ quarks in the final state can be described in different schemes: 
if the hard scale $Q$ of the process is comparable with the bottom-quark mass $m_b$, then the so-called four-flavour scheme (4FS) can be employed, where the heavy quarks are produced in the hard interaction. In other words,  bottom quarks are not present inside the proton, thus the corresponding partonic density is zero. 
On the other hand, if the process is characterised by scales $Q\gg m_b$, one effectively treats the bottom quark as massless and introduces a  five-flavour scheme (5FS). In this case the bottom-quark partonic density inside the proton is perturbatively generated for scales $Q>m_b$. Both schemes have advantages and disadvantages,\footnote{A large literature on this topic exists. See refs.~\cite{Maltoni:2012pa, Lim:2016wjo} for a recent appraisal of the relations and applicability of the schemes and a more comprehensive list of references and ref.~\cite{deFlorian:2016spz} for a thorough presentation of the phenomenological implications and comparisons in searches for additional Higgs bosons in association with $b$ quarks.} thus a careful study to  compare the results in the two cases is required.

As mentioned above, in our study we employ the {\sc DMSimp} UFO models~\cite{Mattelaer:2015haa,Backovic:2015soa,Neubert:2015fka}, which include all ingredients needed in order to perform NLO-accurate predictions in {\sc MG5aMC}.
The models so far have been implemented either with 4FS or 5FS, with corresponding on-shell scheme for the Yukawa's in the case of scalar mediators. In this work, we present the model featuring a scalar mediator, where the $\overline{\textrm{MS}}$ scheme is used for the renormalisation
of the bottom-quark Yukawa coupling (while for the top-quark Yukawa we stick to the on-shell scheme). 
The corresponding renormalisation counterterms read:
\begin{align}
    \delta y_t & = - y_t \frac{g_s^2}{12 \pi^2} \left(\frac{3}{\bar \epsilon} + 4 - 6 \log\frac{m_t}{\mu_R}\right) \,, \\
    \delta y_b & = - y_b \frac{g_s^2}{4 \pi^2\bar \epsilon} \,. 
\end{align}
The $\overline{\textrm{MS}}$ renormalisation scheme allows to automatically resum possibly large logs of the type $\mu_R/m_b$ where the renormalisation scale $\mu_R$ has to be chosen of the order of the mediator mass and introduces the scale dependence for $y_b$. 
In order to take such a dependence into account, notably for the estimate of missing higher-order uncertainties via scale variations, some changes in the code generated by 
{\sc MG5aMC} are necessary, discussed in~\cite{Wiesemann:2014ioa, Degrande:2015vpa, Degrande:2016hyf, Deutschmann:2018avk}.

With the {\sc DMSimp} notation,
the relevant Lagrangian for the DM+\bbbar\ production, i.e.
the interactions of a spin-0 mediator ($Y_0$) with Dirac-type DM candidates ($X_D$) and quarks,
is given by~\cite{Backovic:2015soa}
\begin{align}
 {\cal L} &= \bar X_D (g_{X_D}^S +i g_{X_D}^P\gamma_5)X_D\,Y_0
 +\sum_{q=t,b}\bar q \frac{y_q}{\sqrt{2}} (g_{q}^S +i g_{q}^P\gamma_5)q\,Y_0 \,,
\end{align}  
where $g^{S/P}$ are the scalar/pseudoscalar couplings of DM and quarks.
We normalise the couplings between a spin-0 mediator and quarks to the SM Yukawa couplings,
and explicitly write the Lagrangian only for the third generation quarks.
We note that the top Yukawa indirectly plays a role in DM+\bbbar\ production via the mediator width.
The pure scalar mediator scenario ($Y_0=\phi$) is given by 
$g_{X_D}^S\ne 0$, $g_{q}^S\ne 0$ and $g_{X_D}^P=g_q^P=0$, while
the pure pseudoscalar mediator scenario ($Y_0=a$) is given by 
$g_{X_D}^P\ne 0$, $g_{q}^P\ne 0$ and $g_{X_D}^S=g_q^S=0$.
In this work, we consider such a pure scalar or pseudoscalar scenario, and 
take unity for the coupling parameters as a benchmark.  
In the following, we consider the 13~TeV LHC, and
present the results for both scalar and pseudoscalar mediators
for mass assumptions that range from $10$ GeV to $1$ TeV. 
We assume that the DM mass $m_\chi$ is always $1$ GeV.
The results are valid as long as the condition $m_{\phi/a} > 2 m_{\chi}$ is fulfilled. 

We employ the NNPDF3.0 set, consistent with the order of the computation and with the flavour scheme (the {\tt LHAPDF6}~\cite{Buckley:2014ana} IDs of the LO and NLO 4FS sets are
263400 and 260400 respectively, while the 5FS ones are 263000 and 260000).
The renormalisation and factorisation scales are set to $H_T/3$ ($H_T$ denotes the sum of transverse masses of final-state particles); such a choice
is motivated by the findings of Refs.~\cite{Maltoni:2012pa,Lim:2016wjo}. We stress that this is not the default choice for LO 
(i.e. without the {\sc [QCD]} tag)  and NLO (with the {\sc [QCD]} tag) runs in 
{\sc MadGraph5\_aMC@NLO}. For the former, the default scale is the transverse
mass of the $2\rightarrow2$ partonic system resulting from a
$k_{\mathrm T}$-clustering of final-state partons; for the latter, it is $H_T/2$.  
In all results, we require to have at least one $b$-tagged 
jet with $p_T > 30$\GeV, 
where the jet is defined through the anti-$k_T$ algorithm~\cite{Cacciari:2008gp}, as implemented in {\tt FastJet}~\cite{Cacciari:2005hq,Cacciari:2011ma}, with $R=0.4$.

The generation of the processes in the 4FS can be performed by issuing the following commands:
\begin{verbatim}
import model DMsimp_s_spin0_4f_ybMSbar
generate p p > b b~ xd xd~ / a z w+ w- [QCD]
\end{verbatim}
In the 5FS  in order to have a NLO-accurate description of the one $b$-jet bin, the following commands are necessary:
\begin{verbatim}
import model DMsimp_s_spin0_5f_ybMSbar
generate p p > j xd xd~ [QCD]
\end{verbatim}
Although we only present results of the fixed-order calculations, i.e.\ 
without parton shower, in this note, 
merging of different jet multiplicities in the 5FS at NLO 
can be performed with the FxFx~\cite{Frederix:2012ps} or UNLOPS~\cite{Lonnblad:2012ix} method. 
On the other hand, merging different multiplicities in a 4FS needs special care and cannot be done automatically, yet.

\section{Results}

\begin{figure}
\centering
\begin{subfigure}{.495\textwidth}\centering
\includegraphics[width=\textwidth]{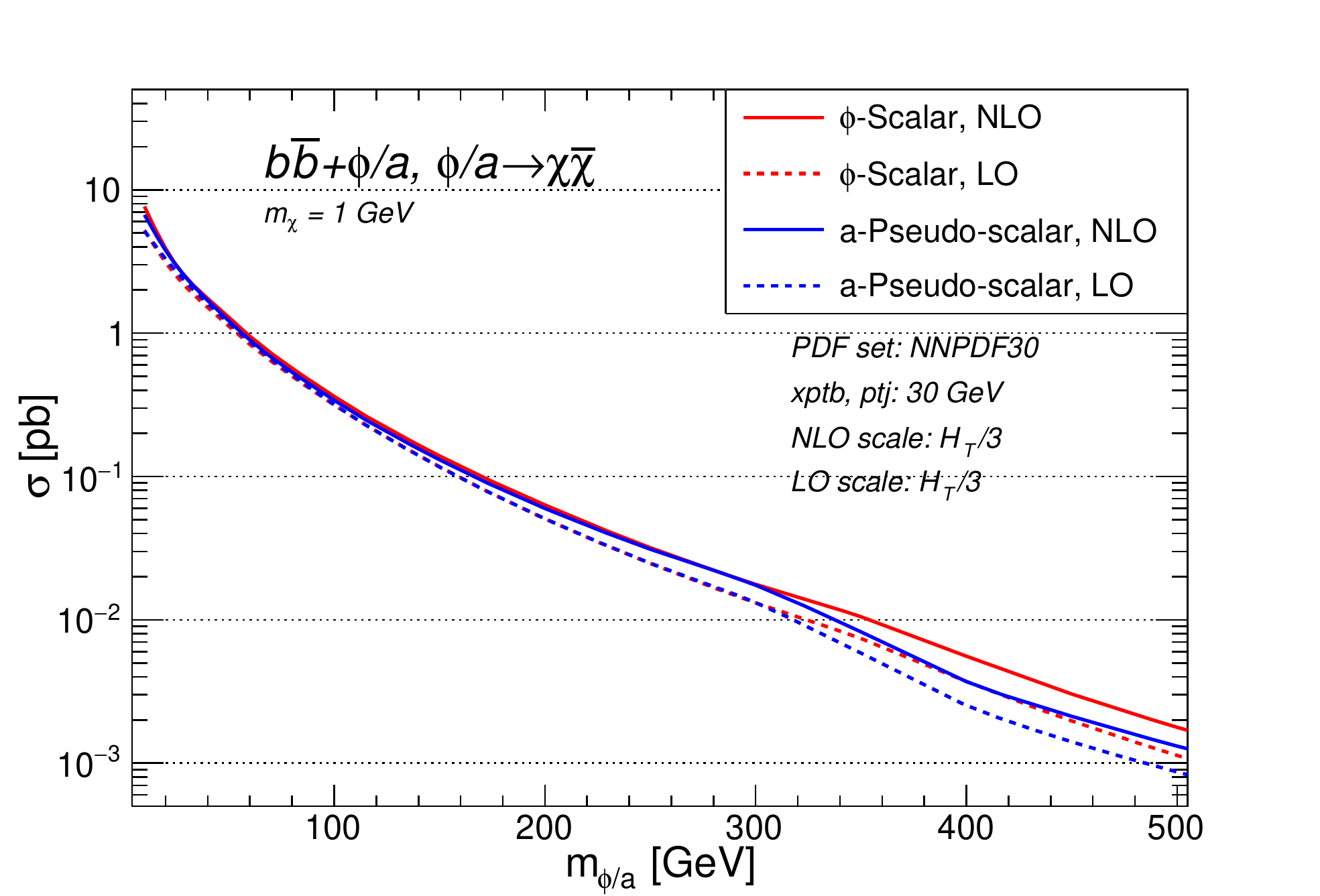}
\caption{}
\label{fig:DMbbxsec1}
\end{subfigure}
\begin{subfigure}{.495\textwidth}\centering
\includegraphics[width=\textwidth]{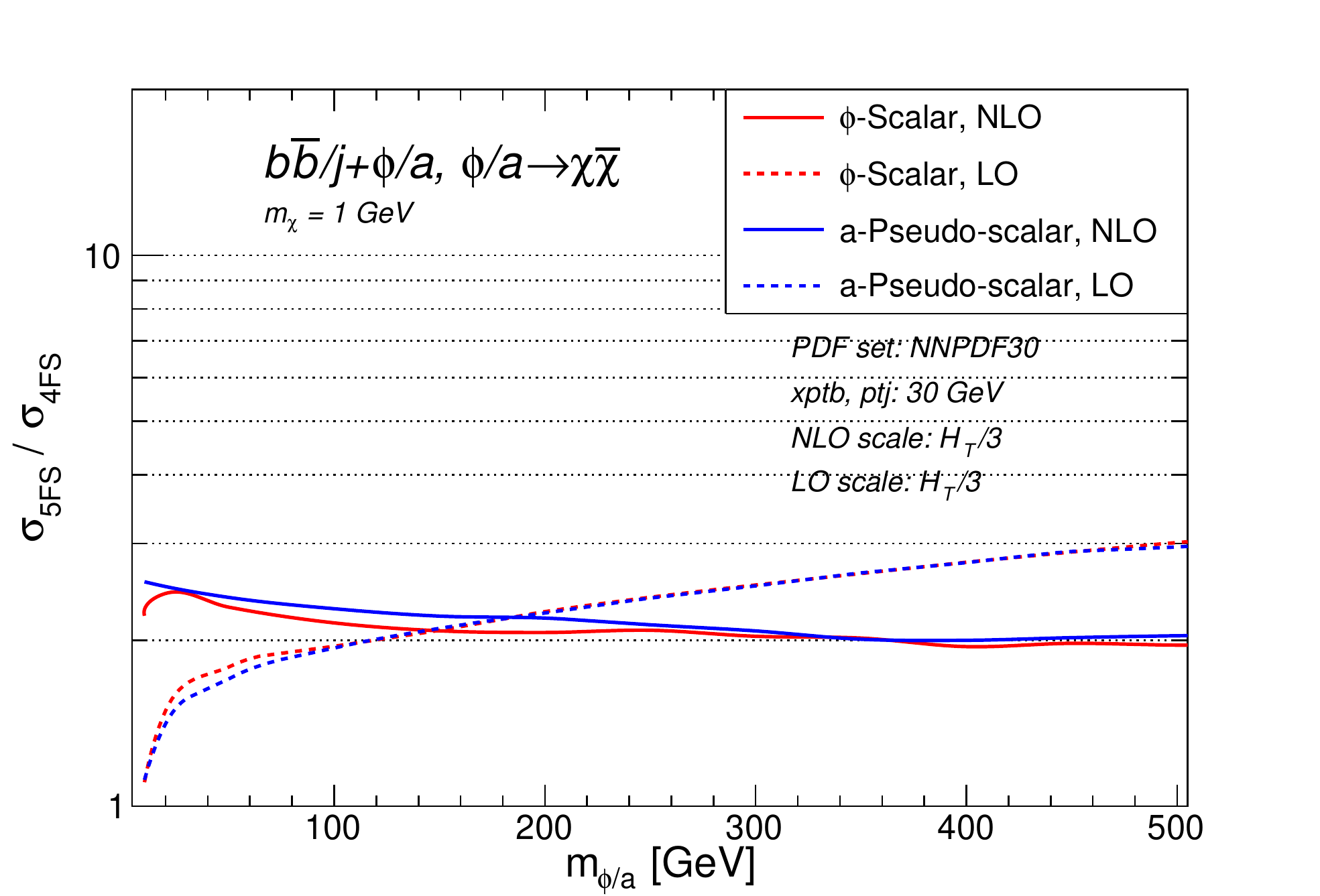}
\caption{}
\label{fig:DMbbxsec2}
\end{subfigure}
\caption{(a) Cross section for
$pp\to b\bar b+\chi\bar\chi$ in the scalar or pseudoscalar mediator models
 at LO and NLO using the 4FS. (b) Ratio of the cross sections calculated in the 4FS and 5FS with a jet \pt\ requirement of 30 GeV.
 }
\label{fig:DMbbxsec}
\end{figure}

Figure~\ref{fig:DMbbxsec1} shows the LO and NLO cross sections for 
$pp\to b\bar b+\chi\bar\chi$
 as a function of the  scalar or
pseudoscalar mediator mass. The DM mass is assumed to be $1\;\GeV$. 
Figure~\ref{fig:DMbbxsec2} shows  the ratio of the cross-section for 5FS and 4FS, for different accuracy orders, with a $b$-jet $p_T$ cut of 30 GeV (referred to as {\tt  xptb} in the figures and captions). The corresponding numbers are quoted in Table~\ref{tab:benchmark_xsecs}. It is found that the different flavour schemes differ by a factor of 2 at NLO-accuracy, which is relatively independent of the mediator mass. On the other hand,  the cross section at LO differs between the two flavour schemes, with a strong dependence on the mediator mass. 
Such a large discrepancy at NLO and for a not-so-exclusive 
observable is quite surprising, and was not observed in Ref.~\cite{Wiesemann:2014ioa}. However, the 5FS simulation employed had only a LO-accurate description of the $b$-jet related observables. In a more recent study~\cite{Krauss:2016orf}, 
where higher multiplicities were included also in the 5FS via NLO merging, a similar discrepancy (factor $\sim 2$) between the 5FS and 4FS was also observed in the one b-jet bin.

\begin{figure}
\centering
\includegraphics[width=.495\textwidth]{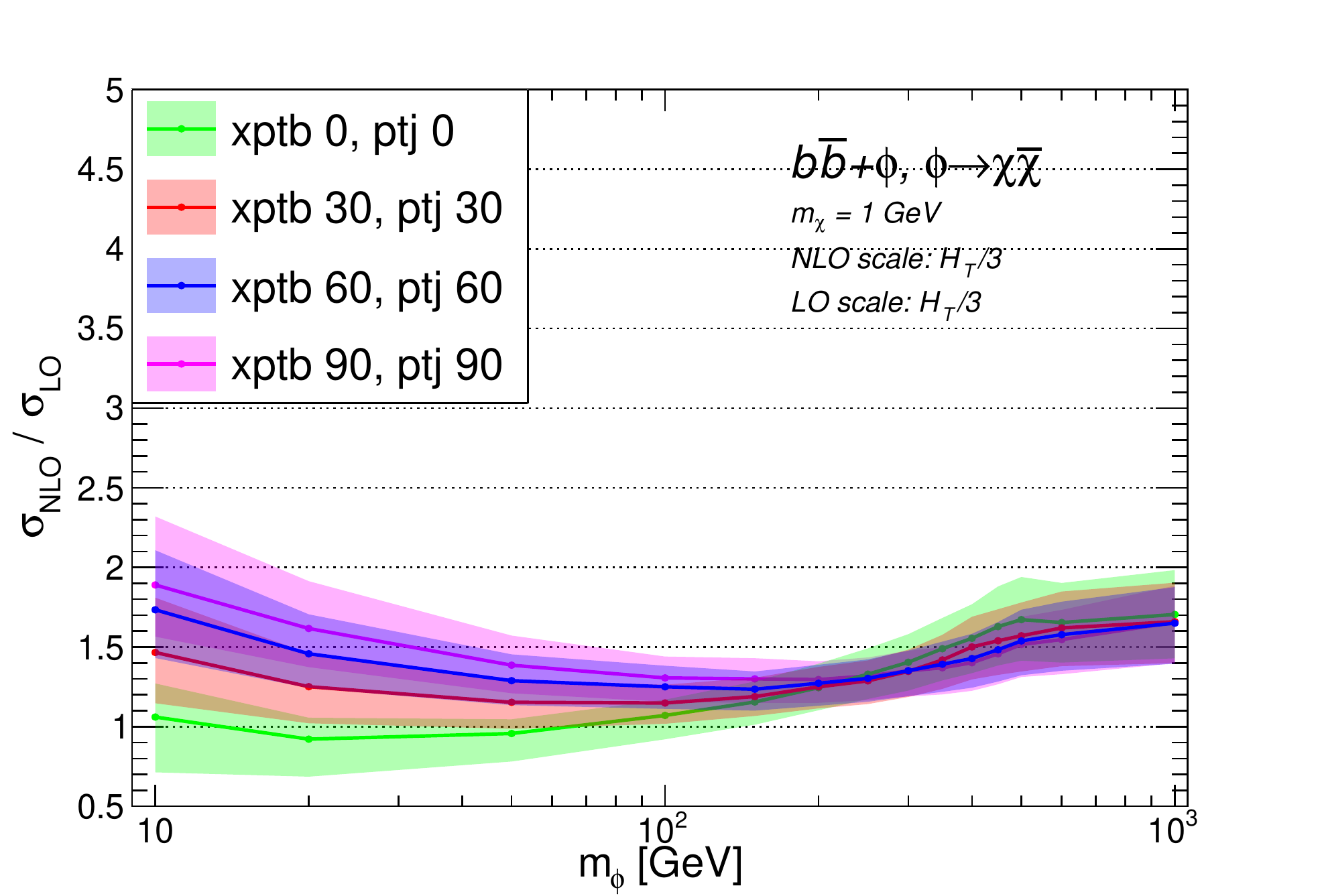}
\includegraphics[width=.495\textwidth]{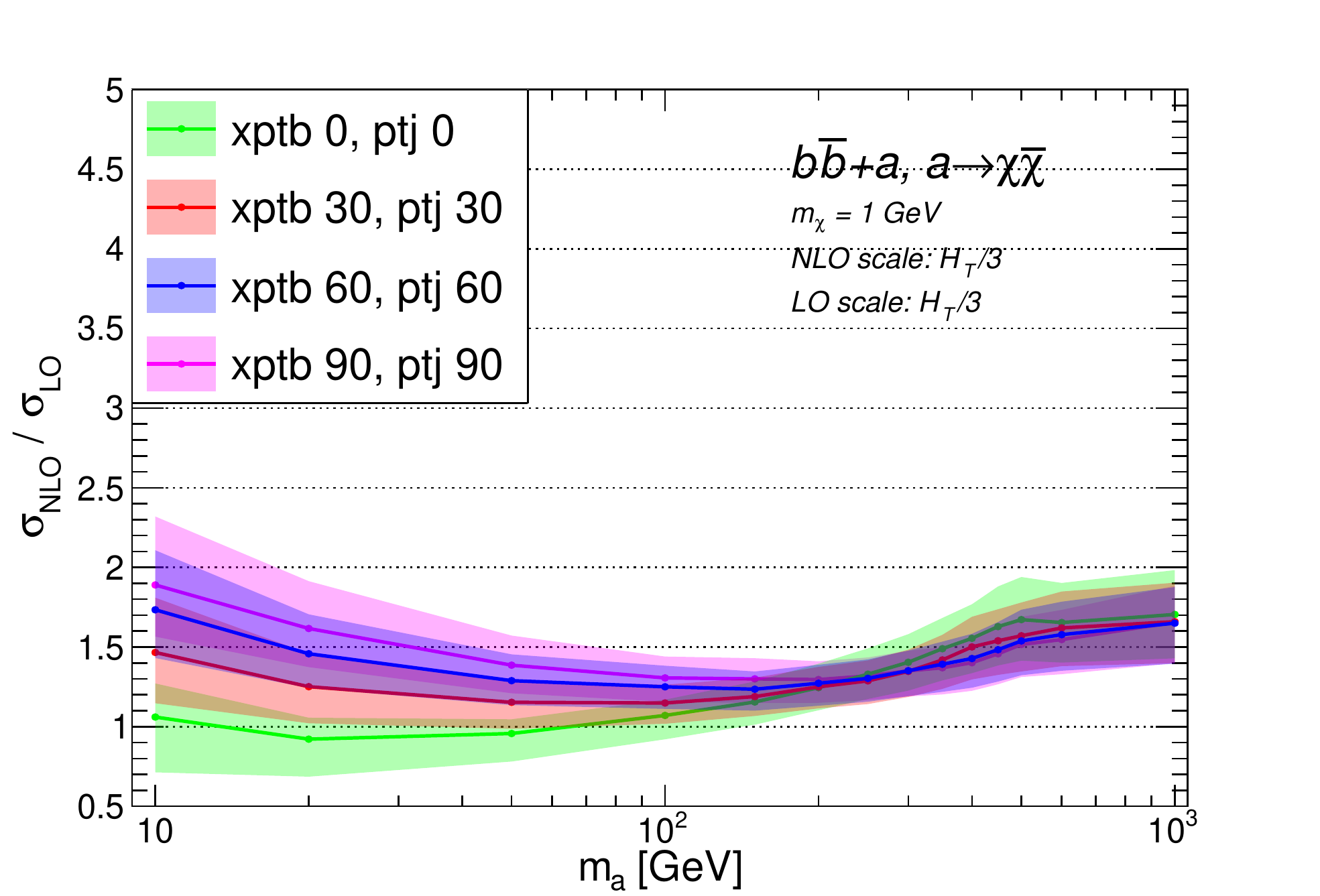}
\caption{Dependence of the NLO/LO K-factors on the {\tt xptb} requirement in the 4FS The uncertainty band around each line includes scales and intra-PDF variations on the NLO cross-section. Scalar mediator models are shown in the upper panel and pseudoscalar mediator models are shown in the bottom panel.}
\label{fig:NLO_bb_xptb}
\end{figure}

The NLO to LO ratio (K factor) in the 4FS is shown in Figure~\ref{fig:NLO_bb_xptb}, for different values of the jet $p_T$ ({\tt xptb}) cut. We observed how a
harder cut turns into larger NLO corrections, at least for mediator masses below 100 GeV. Above this value the K factor is rather insensitive on the 
{\tt xptb} cut.

We conclude by stressing that some extra care should be employed in the definition of the DM$+{b\bar b}$ signal, as it may receive a sizeable contribution from non $y_b$-induced terms. In the case of the SM Higgs boson this has been studied in Ref.~\cite{Deutschmann:2018avk}, where
it has been shown that the $y_t$-induced contribution to $H b \bar b$ is not only larger than is $y_b$-induced counterpart, but it also receives large
NLO corrections. This can be particularly relevant e.g. in two-Higgs-doublet model scenarios for low and intermediate $\tan\beta$.

\begin{table}
\begin{center}

\begin{tabular}{ l  l  l  l  l   }
\hline
$m_{\phi}$ [GeV] &  $\sigma_{NLO,4FS}$ ($H_T/3$, $\overline{\textrm{MS}}$) [pb] & & $\sigma_{NLO,5FS}$ ($H_T/3$, $\overline{\textrm{MS}}$) [pb] & \\
\hline
10                  &             $7.65       \pm      0.06$ & $ ^{+23.4\% + 1.4\%}_{-21.7\% - 1.4\%}$       &      $16.96      \pm       0.15$ & $ ^{+40.1\% + 1.9\%}_{-35.7\% - 1.9\%}$ \\
100                &             $0.36     \pm      0.04$ & $ ^{+9.8\% + 1.6\%}_{-11.2\% - 1.6\%}$     &       $0.779     \pm      0.006$ & $ ^{+5.2\% + 1.9\%}_{-15.1\% - 1.9\%}$ \\
350                  &         $(10.50     \pm     0.10)\cdot 10^{-3}$ & $ ^{+10.9\% + 2.1\%}_{-12.6\% - 2.1\%}$    &       $(21.21     \pm    0.15)\cdot 10^{-3}$ & $ ^{+3.0\% + 2.3\%}_{-5.7\% - 2.3\%}$ \\
500                   &        $(1.78    \pm     0.02)\cdot 10^{-3}$ & $ ^{+13.0\% + 2.6\%}_{-14.0\% - 2.6\%}$     &     $(3.49     \pm     0.03)\cdot 10^{-3}$ & $ ^{+3.3\% + 2.8\%}_{-3.4\% - 2.8\%}$ \\
\hline
\hline
$m_{a}$ [GeV] &  $\sigma_{NLO,4FS}$ ($H_T/3$, $\overline{\textrm{MS}}$) [pb] & & $\sigma_{NLO,5FS}$ ($H_T/3$, $\overline{\textrm{MS}}$) [pb] & \\
\hline
10                 &            $6.69        \pm      0.07$ & $ ^{+18.6\% + 1.4\%}_{-19.7\% - 1.4\%}$       &      $17.10       \pm      0.12$ & $ ^{+40.1\% + 1.9\%}_{-35.7\% - 1.9\%}$ \\
100                &          $0.34      \pm     0.03$ & $ ^{+6.5\% + 1.7\%}_{-9.5\% - 1.7\%}$      &      $0.776     \pm      0.006$ & $ ^{+5.0\% + 1.9\%}_{-15.0\% - 1.9\%}$ \\
350                 &         $(8.23      \pm     0.04)\cdot 10^{-3}$ & $ ^{+11.2\% + 2.1\%}_{-12.1\% - 2.1\%}$     &      $(16.55     \pm    0.11)\cdot 10^{-3}$ & $ ^{+2.8\% + 2.3\%}_{-5.2\% - 2.3\%}$ \\
500                 &        $(1.31    \pm     0.01)\cdot 10^{-3}$ & $ ^{+12.1\% + 2.5\%}_{-13.5\% - 2.5\%}$     &     $(2.68     \pm     0.02)\cdot 10^{-3}$ & $ ^{+2.9\% + 2.8\%}_{-3.7\% - 2.8\%}$ \\
\hline
\end{tabular}
\end{center}
  \caption{NLO cross sections for benchmark points. The generator statistical uncertainties are the symmetric ones next to the cross section values, while the scale and intra-PDF uncertainties are the asymmetric ones on the right of the cross section values.}
\label{tab:benchmark_xsecs}
\end{table}

\section{Conclusions}
We have presented an updated version of the {\sc DMSimp} model, which includes a mixed renormalisation scheme for the
bottom quark mass and Yukawa coupling: the former is renormalised in the on-shell scheme, while the latter in the $\overline{\rm MS}$ one. This choice
is motivated by several studies in the case of Higgs and bottom quarks associated production, and should be employed in future simulations
of  DM+\bbbar. The DM+\bbbar\ NLO cross sections have been investigated under two
different flavour schemes.  
The dependence of these cross sections has also been studied as a
function of various transverse momentum requirements on the jets and
$b$-jets in the final state.  We note that the DM+\bbbar\ NLO cross sections obtained in this paper are
found to be consistent with those used by the ATLAS Collaboration~\cite{Aaboud:2017rzf}, which are
generated in a 4FS setup with an {\tt xptb} requirement of 30 GeV.  On the other hand,  the CMS Collaboration~\cite{Sirunyan:2017xgm} uses  the same models 5FS LO
cross sections without the {\tt xptb} requirement. Given our results, it could be useful to consider  employing a common scheme between ATLAS and CMS,  at least for the normalisation of the predictions. In this respect, a  pragmatic approach such as the Santander matching scheme \cite{Harlander:2011aa} adopted for Higgs in association with $b$ quarks, could also be employed in the case of DM production.

\section*{Acknowledgements}
The research of Y.~Afik and Y.~Rozen was supported by a grant from the United States-Israel Binational Science Foundation, Jerusalem, Israel, and by a grant from the Israel Science Foundation. The work of M.~Zaro is supported by the Netherlands National Organisation for Scientific Research (NWO). This work was started during the BSM session of the workshop "Physics at TeV Colliders" held in Les Houches in June 2017. We thank the organisers for the stimulating environment of the workshop, and the colleagues taking part in the meetings for useful discussions.

\bibliography{DMSimp}

\providecommand{\href}[2]{#2}\begingroup\raggedright\begin{thebibliography}{10}

\bibitem{Abercrombie:2015wmb}
D.~Abercrombie {\em et.~al.}, \href{http://xxx.lanl.gov/abs/1507.00966}{{\tt
  1507.00966}}.

\bibitem{Aaboud:2017rzf}
M.~Aaboud {\em et.~al.},, {\bf ATLAS} Collaboration {\em Eur. Phys. J.} {\bf
  C78} (2018), no.~1 18, [\href{http://xxx.lanl.gov/abs/1710.11412}{{\tt
  1710.11412}}].

\bibitem{Aaboud:2017aeu}
M.~Aaboud {\em et.~al.},, {\bf ATLAS} Collaboration {\em JHEP} {\bf 06} (2018)
  108, [\href{http://xxx.lanl.gov/abs/1711.11520}{{\tt 1711.11520}}].

\bibitem{ATLAS:2016jaa}
T.~A. collaboration,, {\bf ATLAS} Collaboration.

\bibitem{ATLAS:2016ljb}
T.~A. collaboration,, {\bf ATLAS} Collaboration.

\bibitem{ATLAS:2016xcm}
T.~A. collaboration,, {\bf ATLAS} Collaboration.

\bibitem{ATLAS:2016tsc}
T.~A. collaboration,, {\bf ATLAS} Collaboration.

\bibitem{Sirunyan:2017xgm}
A.~M. Sirunyan {\em et.~al.},, {\bf CMS} Collaboration {\em Eur. Phys. J.} {\bf
  C77} (2017), no.~12 845, [\href{http://xxx.lanl.gov/abs/1706.02581}{{\tt
  1706.02581}}].

\bibitem{CMS:2016mxc}
C.~Collaboration,, {\bf CMS} Collaboration.

\bibitem{CMS:2016jxd}
C.~Collaboration,, {\bf CMS} Collaboration.

\bibitem{CMS:2016uxr}
C.~Collaboration,, {\bf CMS} Collaboration.

\bibitem{Alloul:2013bka}
A.~Alloul, N.~D. Christensen, C.~Degrande, C.~Duhr, and B.~Fuks, {\em Comput.
  Phys. Commun.} {\bf 185} (2014) 2250--2300,
  [\href{http://xxx.lanl.gov/abs/1310.1921}{{\tt 1310.1921}}].

\bibitem{Alwall:2014hca}
J.~Alwall, R.~Frederix, S.~Frixione, V.~Hirschi, F.~Maltoni, O.~Mattelaer,
  H.~S. Shao, T.~Stelzer, P.~Torrielli, and M.~Zaro, {\em JHEP} {\bf 07} (2014)
  079, [\href{http://xxx.lanl.gov/abs/1405.0301}{{\tt 1405.0301}}].

\bibitem{Mattelaer:2015haa}
O.~Mattelaer and E.~Vryonidou, {\em Eur. Phys. J.} {\bf C75} (2015), no.~9 436,
  [\href{http://xxx.lanl.gov/abs/1508.00564}{{\tt 1508.00564}}].

\bibitem{Backovic:2015soa}
M.~Backovic, M.~Kr{\"a}mer, F.~Maltoni, A.~Martini, K.~Mawatari, and M.~Pellen,
  {\em Eur. Phys. J.} {\bf C75} (2015), no.~10 482,
  [\href{http://xxx.lanl.gov/abs/1508.05327}{{\tt 1508.05327}}].

\bibitem{Neubert:2015fka}
M.~Neubert, J.~Wang, and C.~Zhang, {\em JHEP} {\bf 02} (2016) 082,
  [\href{http://xxx.lanl.gov/abs/1509.05785}{{\tt 1509.05785}}].

\bibitem{Frederix:2012ps}
R.~Frederix and S.~Frixione, {\em JHEP} {\bf 12} (2012) 061,
  [\href{http://xxx.lanl.gov/abs/1209.6215}{{\tt 1209.6215}}].

\bibitem{deFlorian:2016spz}
D.~de~Florian {\em et.~al.},, {\bf LHC Higgs Cross Section Working Group}
  Collaboration \href{http://xxx.lanl.gov/abs/1610.07922}{{\tt 1610.07922}}.

\bibitem{Maltoni:2012pa}
F.~Maltoni, G.~Ridolfi, and M.~Ubiali, {\em JHEP} {\bf 07} (2012) 022,
  [\href{http://xxx.lanl.gov/abs/1203.6393}{{\tt 1203.6393}}]. [Erratum:
  JHEP04,095(2013)].

\bibitem{Lim:2016wjo}
M.~Lim, F.~Maltoni, G.~Ridolfi, and M.~Ubiali, {\em JHEP} {\bf 09} (2016) 132,
  [\href{http://xxx.lanl.gov/abs/1605.09411}{{\tt 1605.09411}}].

\bibitem{Wiesemann:2014ioa}
M.~Wiesemann, R.~Frederix, S.~Frixione, V.~Hirschi, F.~Maltoni, and
  P.~Torrielli, {\em JHEP} {\bf 02} (2015) 132,
  [\href{http://xxx.lanl.gov/abs/1409.5301}{{\tt 1409.5301}}].

\bibitem{Degrande:2015vpa}
C.~Degrande, M.~Ubiali, M.~Wiesemann, and M.~Zaro, {\em JHEP} {\bf 10} (2015)
  145, [\href{http://xxx.lanl.gov/abs/1507.02549}{{\tt 1507.02549}}].

\bibitem{Degrande:2016hyf}
C.~Degrande, R.~Frederix, V.~Hirschi, M.~Ubiali, M.~Wiesemann, and M.~Zaro,
  {\em Phys. Lett.} {\bf B772} (2017) 87--92,
  [\href{http://xxx.lanl.gov/abs/1607.05291}{{\tt 1607.05291}}].

\bibitem{Deutschmann:2018avk}
N.~Deutschmann, F.~Maltoni, M.~Wiesemann, and M.~Zaro, {\em Submitted to: JHEP}
  (2018) [\href{http://xxx.lanl.gov/abs/1808.01660}{{\tt 1808.01660}}].

\bibitem{Buckley:2014ana}
A.~Buckley, J.~Ferrando, S.~Lloyd, K.~Nordstr{\"o}m, B.~Page, M.~R{\"u}fenacht,
  M.~Sch{\"o}nherr, and G.~Watt, {\em Eur. Phys. J.} {\bf C75} (2015) 132,
  [\href{http://xxx.lanl.gov/abs/1412.7420}{{\tt 1412.7420}}].

\bibitem{Cacciari:2008gp}
M.~Cacciari, G.~P. Salam, and G.~Soyez, {\em JHEP} {\bf 04} (2008) 063,
  [\href{http://xxx.lanl.gov/abs/0802.1189}{{\tt 0802.1189}}].

\bibitem{Cacciari:2005hq}
M.~Cacciari and G.~P. Salam, {\em Phys. Lett.} {\bf B641} (2006) 57--61,
  [\href{http://xxx.lanl.gov/abs/hep-ph/0512210}{{\tt hep-ph/0512210}}].

\bibitem{Cacciari:2011ma}
M.~Cacciari, G.~P. Salam, and G.~Soyez, {\em Eur. Phys. J.} {\bf C72} (2012)
  1896, [\href{http://xxx.lanl.gov/abs/1111.6097}{{\tt 1111.6097}}].

\bibitem{Lonnblad:2012ix}
L.~L{\"o}nnblad and S.~Prestel, {\em JHEP} {\bf 03} (2013) 166,
  [\href{http://xxx.lanl.gov/abs/1211.7278}{{\tt 1211.7278}}].

\bibitem{Krauss:2016orf}
F.~Krauss, D.~Napoletano, and S.~Schumann, {\em Phys. Rev.} {\bf D95} (2017),
  no.~3 036012, [\href{http://xxx.lanl.gov/abs/1612.04640}{{\tt 1612.04640}}].

\bibitem{Harlander:2011aa}
R.~Harlander, M.~Kr{\"a}mer, and M.~Schumacher,
  \href{http://xxx.lanl.gov/abs/1112.3478}{{\tt 1112.3478}}.

\end{thebibliography}\endgroup

\end{document}